# NRQCD and Static Systems – A General Variational Approach [*]

Terrence Draper, Craig McNeile and Constantine Nenkov [a]

[a]Department of Physics and Astronomy, University of Kentucky, Lexington, KY 40506, USA

We present initial results from Monte Carlo simulations of NRQCD-light, static-light, and NRQCD-NRQCD mesons, using a variational technique (MOST), as part of our ongoing calculation of the $f_B$ decay constant. The basis states for the variational calculation are quark-antiquark operators separated by all possible relative distances not equivalent under the cubic group (for example, for a $20^3$ lattice there are 286 operators). The efficacy of the method is demonstrated by the good plateaus obtained for the ground state and the clean extraction of the wave functions of the ground and first radially excited state.

Theoretical uncertainties are starting to dominate the determination from experiment of the Cabibbo-Kobayashi-Maskawa quark mixing matrix, with a major bottleneck being the reliable calculation of the heavy-light pseudoscalar decay constant, $f_B$ [1]. An essential part of its determination on the lattice is its value in the static [2] and nonrelativistic (NRQCD) [3] approximations. Unfortunately, the static approximation is plagued by the particularly bad signal-to-noise ratio for the correlation functions. The use of several standard lattice techniques has led to differing results from various groups in reporting values of the heavy-light pseudoscalar decay constant in the static approximation on the lattice [4]. This can be traced to the difficulty in isolating unambiguously the ground state at early Euclidean time separations [5].

The FNAL group were able to get reliable results for $f_B^{\text{stat}}$ [6] from a variational calculation using a basis set of trial smearing functions designed to have good overlap with the physical states, having been constructed as wave functions of a simple relativistic quark model Hamiltonian.

The present authors used a complementary and more general approach [7]: "MOST" (Maximal Operator Smearing Technique). With MOST, optimally smeared operators are generated without any ansatz, and emerge naturally and directly from the QCD dynamics. This variational calculation uses a *complete* basis set of $\overline{Q}q$ interpolating operators

$$\Phi_{s_i}(\vec{x},t) = \sum_{\vec{r}} s_i(\vec{r})\overline{Q}(\vec{x}+\vec{r},t)\Gamma q(\vec{x},t) \qquad (1)$$

Each $s_i$ is a sum of delta functions of relative separations $\vec{r}$ equivalent under cubic symmetry. The complete set of these "onion shells" is used to construct the $N \times N$ two-point correlation matrix $C(t)$ (with $N = 165$ and 286 for lattice volumes of $16^3$ and $20^3$ respectively). This matrix contains *all* of the information available from two-quark interpolating fields. It can be analyzed at leisure on a workstation using a variety of different techniques including, for example, multi-exponential fits to a submatrix defined by model wave functions, or an iterative method based on using a measured wave function as a source. The best method will be a compromise between maximizing the information obtained from the matrix and minimizing the effects of the noisy elements. We use a general approach to extract the amplitudes $\mathcal{A}_{in} = \langle \text{vac}|\Phi_{s_i}|n\rangle$ (which are the desired Bethe-Salpeter wave functions). If the wave functions were orthogonal they could be extracted from a straightforward diagonalization of the matrix; however, since they are not, the dual wave functions, $\mathcal{Z}$, are required. These satisfy $\mathcal{Z}^\dagger \mathcal{A} = 1$, and can be obtained by diagonalization of $C(t_0)^{-1/2}C(t)C(t_0)^{-1/2}$ [8,9]. These "$\mathcal{Z}$"-sources – optimally-smeared interpolating oper-ators, are then used to construct local-smeared and smeared-smeared correlators, from which $f_B$ is extracted in the usual way. The optimal interpolating operators project out ground and excited states at early Euclidean times when statistical noise is smallest.

---

[*]Presented by T. Draper at Lattice '94, Bielefeld. This work is supported in part by the U.S. Department of Energy under grant numbers DE-FG05-84ER40154 and DE-FC02-91ER75661, by the National Science Foundation under grant number EHR-9108764 and by the Center for Computational Sciences, University of Kentucky.

The method can be applied to create optimal smearing functions for lattice simulations of any heavy-quark effective theory, whenever it is computationally inexpensive to compute many heavy quark propagators. It is then feasible to use the complete basis, using fast-Fourier transforms to evaluate the convolutions. Here we focus on the application of MOST to heavy-light mesons, where the heavy quark is described by a nonrelativistic action (NRQCD) [3,10], and whose propagation can be computed at low cost, relative to that of a light Wilson quark, by recursion [3]. As a by-product, one can construct NRQCD-NRQCD mesons for free.

Light quark propagators were computed for $\kappa = 0.154$, 0.155, and 0.156 with the Wilson fermionic action in a background of 32 quenched (Coulomb gauge-fixed) configurations (selected every 1000 pseudo-heat-bath sweeps after 5000 thermalization sweeps) with $\beta = 6.0$ on a $20^3 \times 30$ lattice.

Reference [7] contains our preliminary result for $f_B^{\text{stat}}$. See reference [11] for a complete set of results for decay constants, binding energies and wave functions for static-light mesons. For example, for the static-light meson we find

$$\frac{f_{B_s}^{\text{stat}}}{f_{B_u}^{\text{stat}}} = 1.22^{+1}_{-1} \text{ (statistical)} \qquad (2)$$

Here we present a sampling of the comparison of the properties of static-light, NRQCD-light and NRQCD-NRQCD mesons. Figure 1 displays an effective-"mass" plot for the local-smeared channel computed using an optimal ("$\mathcal{Z}$") source designed to project out the ground state of the static-light pseudoscalar meson. Figures 2 and 3 are similar plots for the NRQCD-light meson and the NRQCD-NRQCD meson (bottomonium), respectively. (The values of the effective masses are renormalized differently for different actions and so should not be compared.) In all three cases, MOST is successful in producing effective-mass plots which plateau early, are flat indicating ground-state saturation and have small statistical errors. Figures 4, 5, and 6 show $1S$ and $2S$ Bethe-Salpeter (Coulomb gauge) wave functions (normalized to unity at $\vec{r} = 0$) for static-light, NRQCD-light, and NRQCD-NRQCD mesons, respectively. A simple NRQCD action (tadpole-improved, but with only the kinetic energy term with bare lattice mass $Ma = 1.7$) is used here in this illustrative example. Statistical errors seem to be as small as in the static case, and will permit a precise comparison of the static $f_B$ results with those forthcoming for an NRQCD heavy quark including $1/M$ terms in the action and current [12].

We plan to apply these techniques to simulations using the Mandula-Ogilvie [13] lattice implementation of the Isgur-Wise fixed-velocity action, where there have been problems in extracting a consistent signal using simple smearing methods.

## REFERENCES

1. J. Rosner, *J. Phys. G: Nucl. Part. Phys.* **18**, 1575 (1992), for example.
2. E. Eichten, *Nucl. Phys. B (Proc. Suppl.)* **4**, 170 (1988).
3. B. Thacker and G. Lepage, *Phys. Rev. D* **43**, 196 (1991).
4. C. Bernard, *Nucl. Phys. B (Proc. Suppl.)* **34**, 47 (1994).
5. R. Sommer, these proceedings.
6. A. Duncan *et al.*, Properties of $B$-mesons in lattice QCD, preprint FERMILAB-PUB-94/164-T, 1994.
7. T. Draper and C. McNeile, *Nucl. Phys. B (Proc. Suppl.)* **34**, 453 (1994).
8. M. Lüscher and U. Wolff, *Nucl. Phys.* **339**, 222 (1990).
9. A. Kronfeld, *Nucl. Phys. B (Proc. Suppl.)* **17**, 313 (1990).
10. J. Sloan, these proceedings, and references therein.
11. T. Draper and C. McNeile, in preparation.
12. T. Draper, C. McNeile and C. Nenkov, in preparation.
13. J. Mandula and M. Ogilvie, *Nucl. Phys. B (Proc. Suppl.)* **34**, 480 (1994).

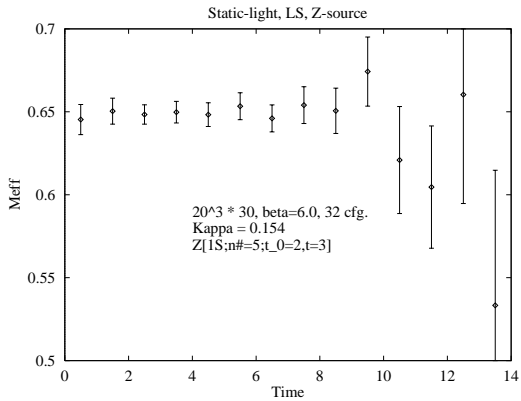

Figure 1. Ground state static-light meson effective mass.

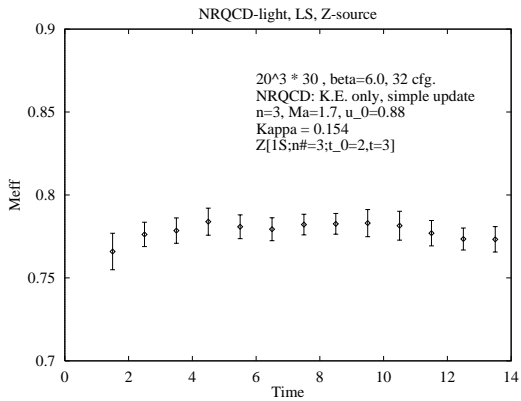

Figure 2. Ground state NRQCD-light meson effective mass.

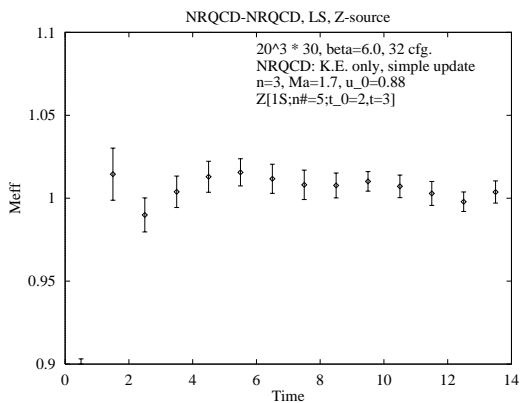

Figure 3. Ground state NRQCD-NRQCD meson effective mass.

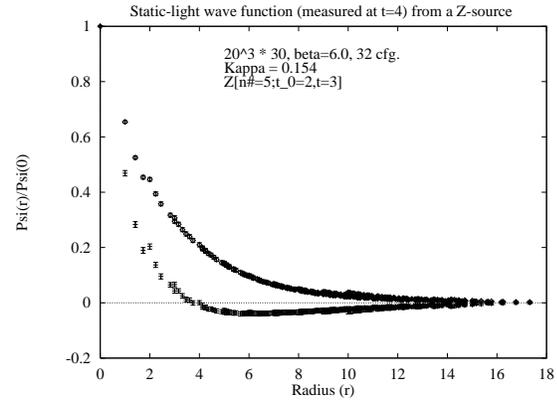

Figure 4. 1S and 2S wave functions normalized to unity at the origin for a static-light meson.

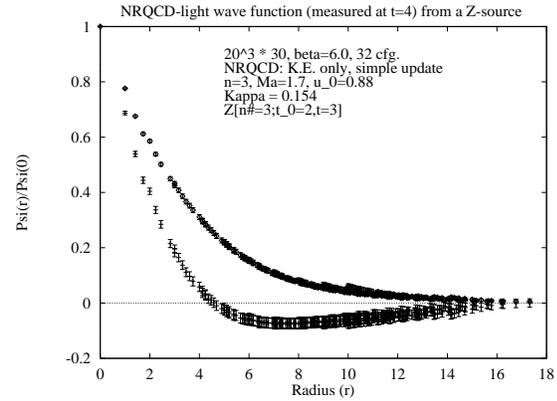

Figure 5. 1S and 2S wave functions normalized to unity at the origin for a NRQCD-light meson.

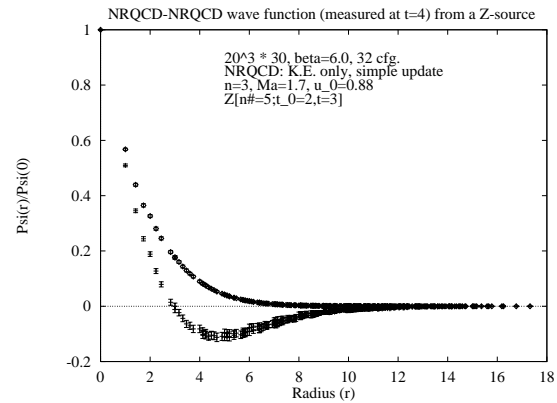

Figure 6. 1S and 2S wave functions normalized to unity at the origin for a NRQCD-NRQCD meson.